\documentclass[a4paper]{jpconf}
\usepackage{graphicx}
\usepackage{caption}
\pdfoutput=1
\begin{document}
\title{Results from Seven Years of AMANDA-II}

\author{Tyce DeYoung, for the IceCube Collaboration}

\address{Department of Physics and Center for Particle Astrophysics,
  Pennsylvania State University, University Park, PA 16802, USA}

\ead{deyoung@phys.psu.edu}


\begin{abstract}
  AMANDA is a first-generation high energy neutrino telescope, which
  has taken data at the South Pole in its final configuration since
  2000.  Results from seven years of operation are presented here,
  including observation of the atmopheric neutrino flux and searches
  for astrophysical neutrinos from cosmic ray accelerators, gamma ray
  bursts, and dark matter annihilations.  In 2007, AMANDA was
  incorporated into the IceCube neutrino telescope, where its higher
  density of instrumentation improves the low energy response.  In the
  near future, AMANDA will be replaced by the IceCube Deep Core, a
  purpose-built low energy extension of IceCube.
\end{abstract}

\section{Introduction}
The Antarctic Muon And Neutrino Detector Array (AMANDA) was
constructed at the Amundsen-Scott South Pole Station over a period of
years, reaching completion in 2001.  The completed detector operated
independently until 2006.  AMANDA's primary goal was the discovery of
high energy neutrinos from astrophysical sources, in the energy range
of roughly 100 GeV to 100 PeV.  As a first-generation instrument,
AMANDA served as a proof of concept and prototype for later neutrino
telescopes such as IceCube.

Neutrinos are detected by AMANDA when they undergo
either charged-current (CC) or neutral-current (NC) interactions with
nucleons in the ice.   Muon neutrino CC interactions
produce a secondary muon which may travel kilometers through the ice.
Neutral-current interactions and CC electron neutrino events produce
hadronic or electromagnetic showers of particles near the interaction
vertex but no long track.  In either case, the secondary particles are
detected via 
Cherenkov radiation by a sparse three-dimensional array of
photosensors embedded in the extremely transparent polar ice cap
(optical attenuation length $\sim 25$ m).

The final detector configuration, known as AMANDA-II, consisted of 677
optical modules (OMs) deployed on 19 
strings at depths of roughly 1500 m -- 2000 m below the surface of the
ice sheet.  Each optical module consisted of an 8'' 
photomultiplier tube (PMT) 
contained within a glass
pressure vessel.  The OM received HV power from the surface, and
returned PMT signals amplified by a gain of $10^8$--$10^9$ to the
surface over either electrical or fiber optic cable.  The strings were
arranged in concentric circles, with an 
outermost radius of 100 m, resulting in an instrumented volume of
approximately 15 Mton of ice.

After seven years of successful operation, AMANDA was incorporated
into the second-generation IceCube neutrino telescope, now half
complete.  Independent operation thus ceased after the 2006 data run,
with a total accumulated exposure of 3.8 years after accounting for
detector maintenance periods and deadtime in the data acquisition system.

\section{Atmospheric Neutrinos}

The flux of atmospheric neutrinos produced by decaying $\pi$ and $K$
mesons in cosmic ray air showers in the Earth's atmosphere constitutes
a background to searches for astrophysical neutrinos, but is also a
useful calibration source for AMANDA. This flux is known to a
precision of about 30\% in the energy range to which AMANDA is
sensitive \cite{Barr:2006it}.  Neutrinos can be
identified within the flux of penetrating cosmic ray muons by
searching for muon tracks originating from below the horizon; these
upgoing muons must have been produced by neutrino interactions.

\begin{figure}[h]
  \begin{minipage}{15pc}
    \includegraphics[height=2.4in]{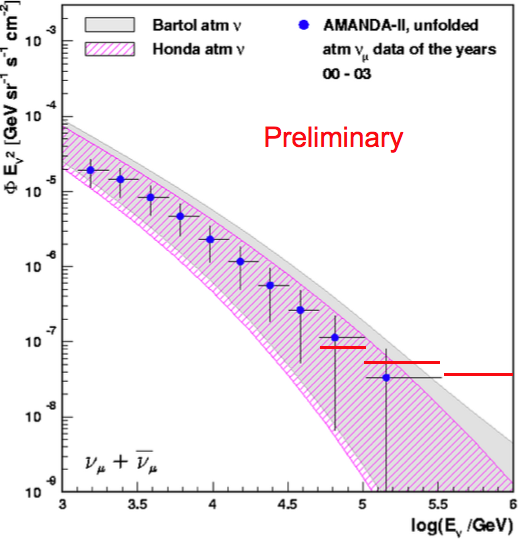}
    \caption{\label{fig:atmnu}Atmospheric muon neutrino flux observed
      by AMANDA.  The red lines indicate a limit placed on a possible
      $E^{-2}$ component.}
  \end{minipage}\hspace{2pc}%
  \begin{minipage}{20pc}
    \centering
    \includegraphics[height=2.4in]{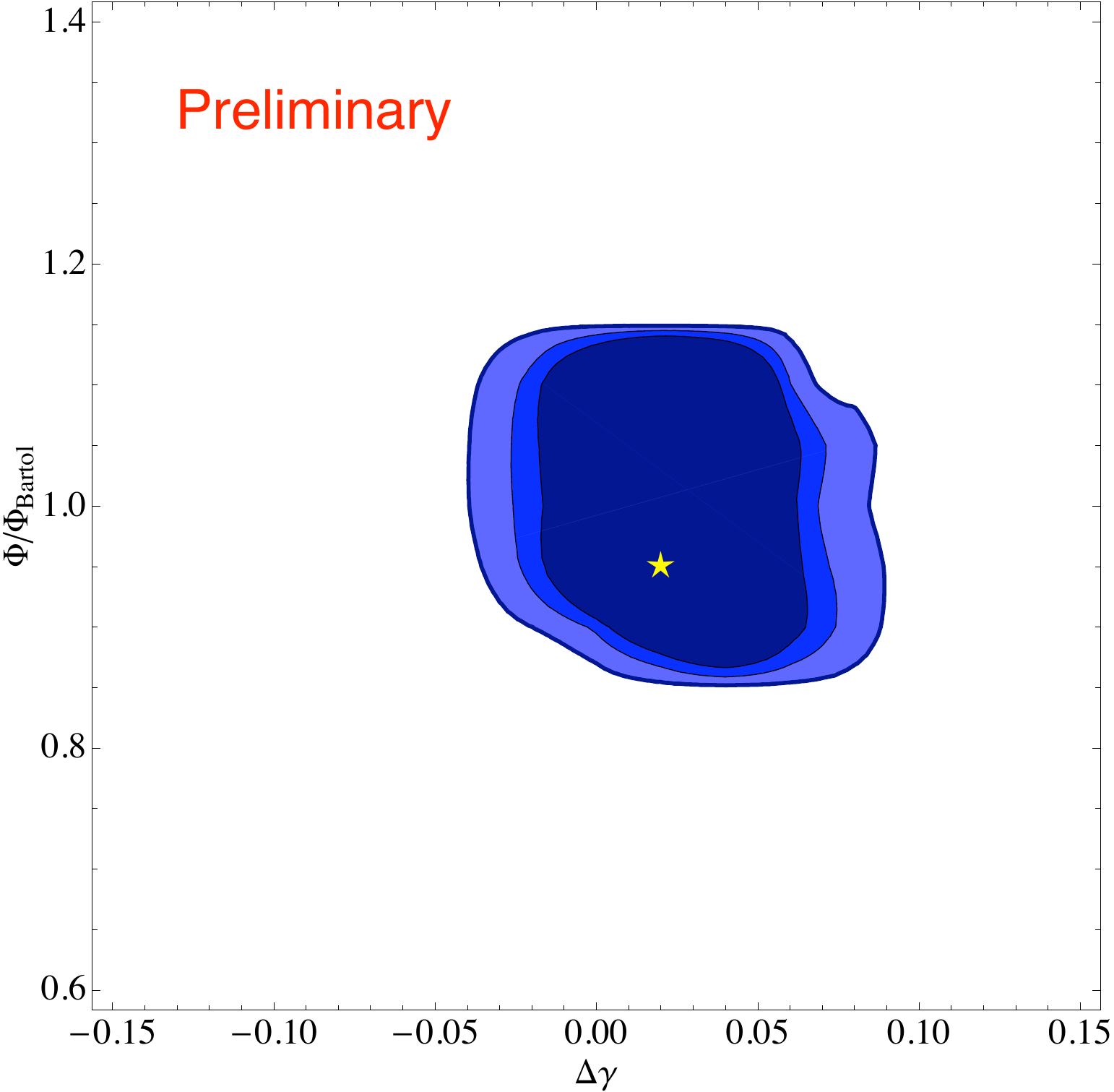}
    \caption{\label{fig:atmnu_7yr}The 90\%, 95\%, and 99\%
      C.L. contours and best fit point in flux 
      ($\Phi$) and spectral index ($\gamma$) of the 
      atmospheric neutrino flux obtained from one simulated
      analysis of the  seven year data set.}  
  \end{minipage} 
\end{figure}

The AMANDA measurement of the atmospheric muon neutrino energy
spectrum is shown in Figure~\ref{fig:atmnu}, using data taken over the
four year period 2000--2003 \cite{munich}. The theoretical bands
include the systematic uncertainties due to our modeling of the
detector and the propagation of light through the ice. Because the
muons that are detected are produced in neutrino interactions at
some distance from the detector, the energy of the incident neutrino
is not directly measured, but must be inferred from the observed
muon energy and the known energy loss rate for muons
traversing ice and rock. The atmospheric neutrino spectrum is thus
statistically unfolded from the observed muon spectrum, leading to the
correlated statistical error bars on the points in
Fig.~\ref{fig:atmnu}. The measurement illustrates that the response of
the detector to muon neutrinos is well understood, although the
precision is insufficient at present to distinguish theoretical models.


The fact that the observed spectrum is consistent with the predictions
for the atmospheric flux at low energies allows us to place a limit on
the magnitude of any possible diffuse muon neutrino flux with a harder
spectrum than the conventional atmospheric flux, for example due to
the prompt decays of charmed leptons in atmospheric air showers or to
a population of unresolved astrophysical sources.  That limit is
indicated with the red bars shown in the highest energy bins of
Fig.~\ref{fig:atmnu}.  Results from the full seven year
data set will be available soon; the expected precision of the
atmospheric flux measurement from that
study, derived from analysis of a single simulated data set based on
the Bartol flux \cite{Barr:2006it}, is shown in
Fig.~\ref{fig:atmnu_7yr}.  

In addition, the very large number of atmospheric neutrinos detected
by AMANDA can be used to search for evidence of physics beyond the
Standard Model, such as quantum decoherence or violations of Lorentz
invariance \cite{btsm}. These phenomena would appear as
oscillation-like behavior, 
but at higher energies than standard neutrino oscillations and with
very different characteristic flavor signatures.  


\section{Astrophysical Neutrinos}

The primary goal of 
AMANDA is
the detection of individual astrophysical sources of  high energy
neutrinos.  We have conducted an unbinned maximum likelihood search
for point sources of neutrinos, using a sky map of 6,595 upward going
neutrino events collected over seven years \cite{7yr}.  The
direction 
of each of these events was reconstructed using a maximum likelihood
method accounting for the propagation of light in the ice
\cite{Ahrens:2003fg}, and the 
width of the solution in the likelihood space was used to
estimate the angular resolution for each event.  

\begin{figure}
  \begin{center}
    \includegraphics[width=0.8\textwidth]{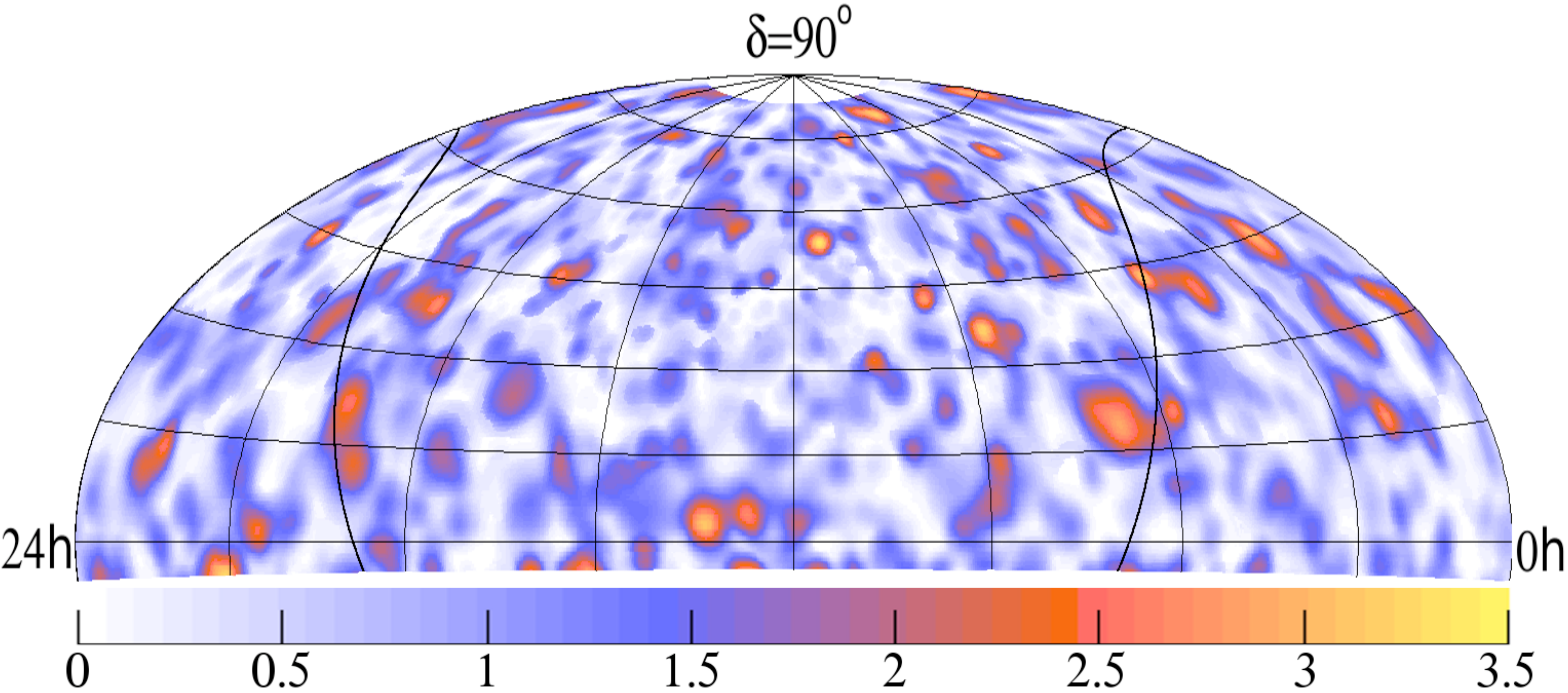}
  \end{center}
  \caption{\label{fig:skymap}Map of the pre-trials significances
    obtained from an unbinned point source search using the full data
    set.  The most significant point has a significance of
    3.38$\sigma$; 95\% of randomized background skymaps include a
    point at least as significant.}
\end{figure}

With these data, the significance of the deviation from a background
uniform in right ascension ($\alpha$) was calculated for all points on
the sky with declinations ($\delta$) between $-5^\circ$ and
$85^\circ$.  The results are 
shown in Figure~\ref{fig:skymap}.  The most significant point on the
sky
had a significance of $3.38\sigma$ before accounting for trial
factors.  The true 
significance was assessed by repeatedly randomizing the right
ascension of each event to create pure background maps, with any
possible real sources smeared out across the sky.  In 95\% of such
maps, a point with at least $3.38\sigma$ was found, indicating
that such a level is 
consistent with statistical fluctuations of the background.

\begin{figure}
  \begin{minipage}{19pc}
    \begin{center}
      \includegraphics[height=2.4in]{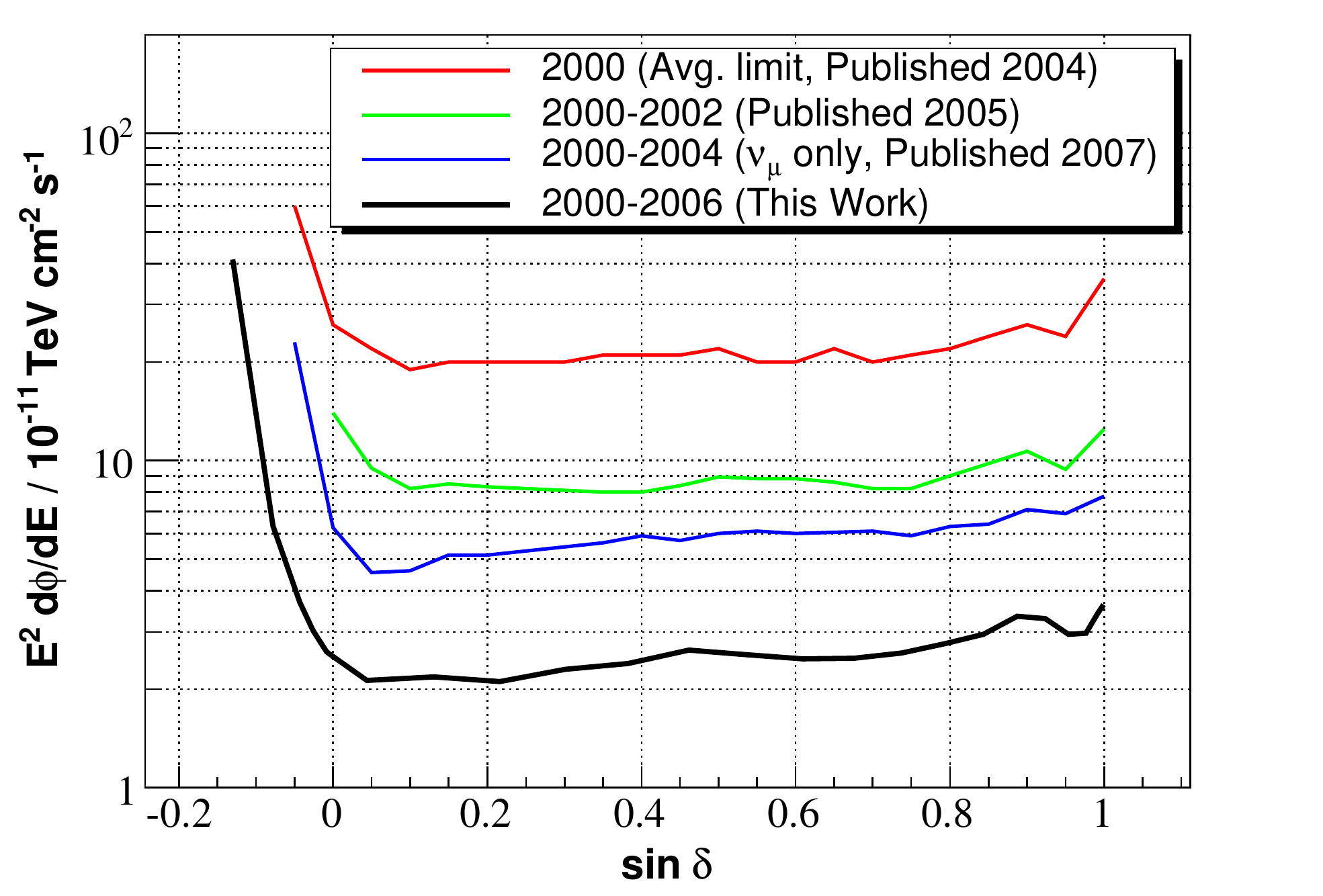}
      \caption{\label{fig:ptsrcsens}Average 90\% C.L. upper limit on
        the $\nu_\mu + \bar{\nu}_\mu$ 
        flux (assuming flavor equality) from the unbinned point
        source search as a function of declination, compared to the
        results of previous searches using portions of the full data
        set.} 
    \end{center}
  \end{minipage}
  \hspace{3pc}%
  \begin{minipage}{15pc}
    \begin{center}
      \lineup
      \begin{tabular}{*{3}{l}}
        \br            
        Source & $\Phi_{90}$ & $p$-value \cr
        \mr
        Crab Nebula   & 9.27 & 0.10 \cr 
        MGRO J2019+37 & 9.67 & 0.077 \cr
        Mrk 421       & 2.54 & 0.82 \cr
        Mrk 501       & 7.28 & 0.22 \cr
        LS I +61 303  & 14.74 & 0.034 \cr
        Geminga       & 12.77 & 0.0086 \cr
        1ES 1959+650  & 6.76 & 0.44 \cr
        M87           & 4.49 & 0.43 \cr
        Cygnus X-1    & 4.00 & 0.57 \cr
        \br
      \end{tabular}
      \vspace{0.1in}
      \captionof{table}{\label{tab:sourcelimits}Selected results from
        a search for neutrino emission from 26 predefined source
        candidates
        (details in text). 
        The probability of 
        $p \leq 0.0086$ for at least one of 26 sources is 20\% in the
        absence of signal.}
    \end{center}
  \end{minipage} 
\end{figure}

The sensitivity of this search to point sources of neutrinos is shown
in Figure~\ref{fig:ptsrcsens}, compared to the sensitivities of
previous AMANDA analyses
\cite{Ahrens:2003pv,Ackermann:2004aga,Achterberg:2006vc}.   The
increase in sensitivity is due to 
improvements in analysis technique as well as to the increased data
collected, with a total improvement of an order of magnitude over the
initial limit based on one year of data.


In addition to the unbinned search for point sources of neutrinos
anywhere in the Northern sky, we have also searched for emission from
a list of 26 candidate sources selected {\it a priori} on theoretical
grounds \cite{7yr}.  Results for several of the sources are given in
Table~\ref{tab:sourcelimits}, including the source with the lowest
chance probability ($p$-value), Geminga.  The 90\% C.L. upper limits
on muon and tau neutrinos are given as $E^2 \Phi_{\nu_\mu + \nu_\tau}
< \Phi_{90} \times 10^{-11}$ TeV cm$^{-2}$ 
s$^{-1}$ assuming flavor equality, and the $p$-values do not take into
account the  
number of statistical trials.  No evidence for neutrino emission from
these sources has been observed.  

\begin{figure}[b]
  \begin{center}
    \includegraphics[height=2.7in]{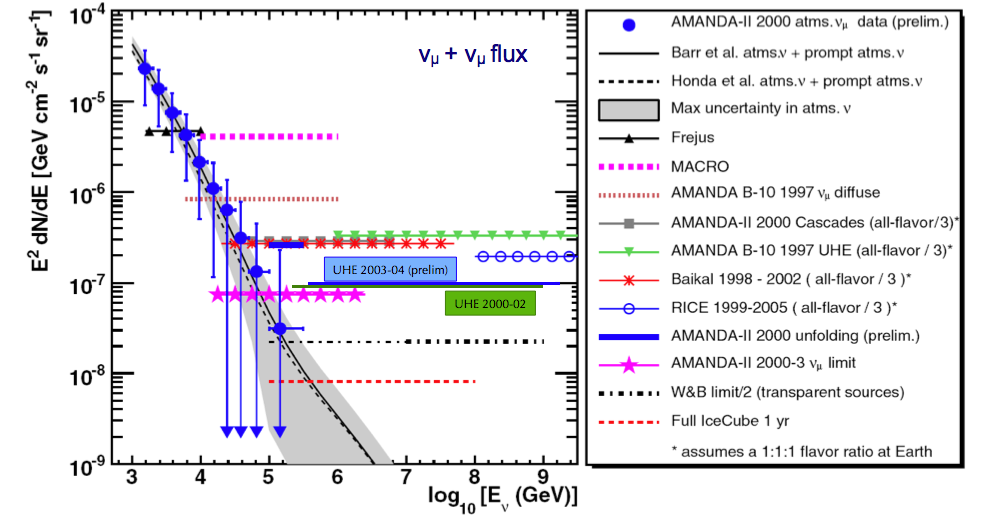}
  \end{center}
  \caption{\label{fig:diffuse}Integral limits on possible $E^{-2}$
    diffuse fluxes.  The energy ranges shown are those  which produce the
    central 90\% of events in each analysis.}
\end{figure}

Searches for diffuse fluxes of astrophysical neutrinos have been
carried out as well
\cite{ultrahigh,Achterberg:2007qp,Ackermann:2005sb}.  Such diffuse
astrophysical fluxes are 
generically predicted to have harder spectra than the atmospheric
neutrinos, so these analyses search for evidence of higher energy events
than would be expected from the atmospheric flux.  The limits obtained
by these searches are shown in Fig.~\ref{fig:diffuse}.  Flavor
equality due to complete mixing over cosmological baselines has been
assumed, and the flux refers to the muon component alone.  Present
limits are within an order of magnitude of the Waxman-Bahcall level
\cite{Waxman:1998yy}, 
with IceCube predicted to probe lower fluxes with a year or less of
operation with the complete detector.

\begin{figure}[h]
  \begin{minipage}{18pc}
    \includegraphics[width=17pc]{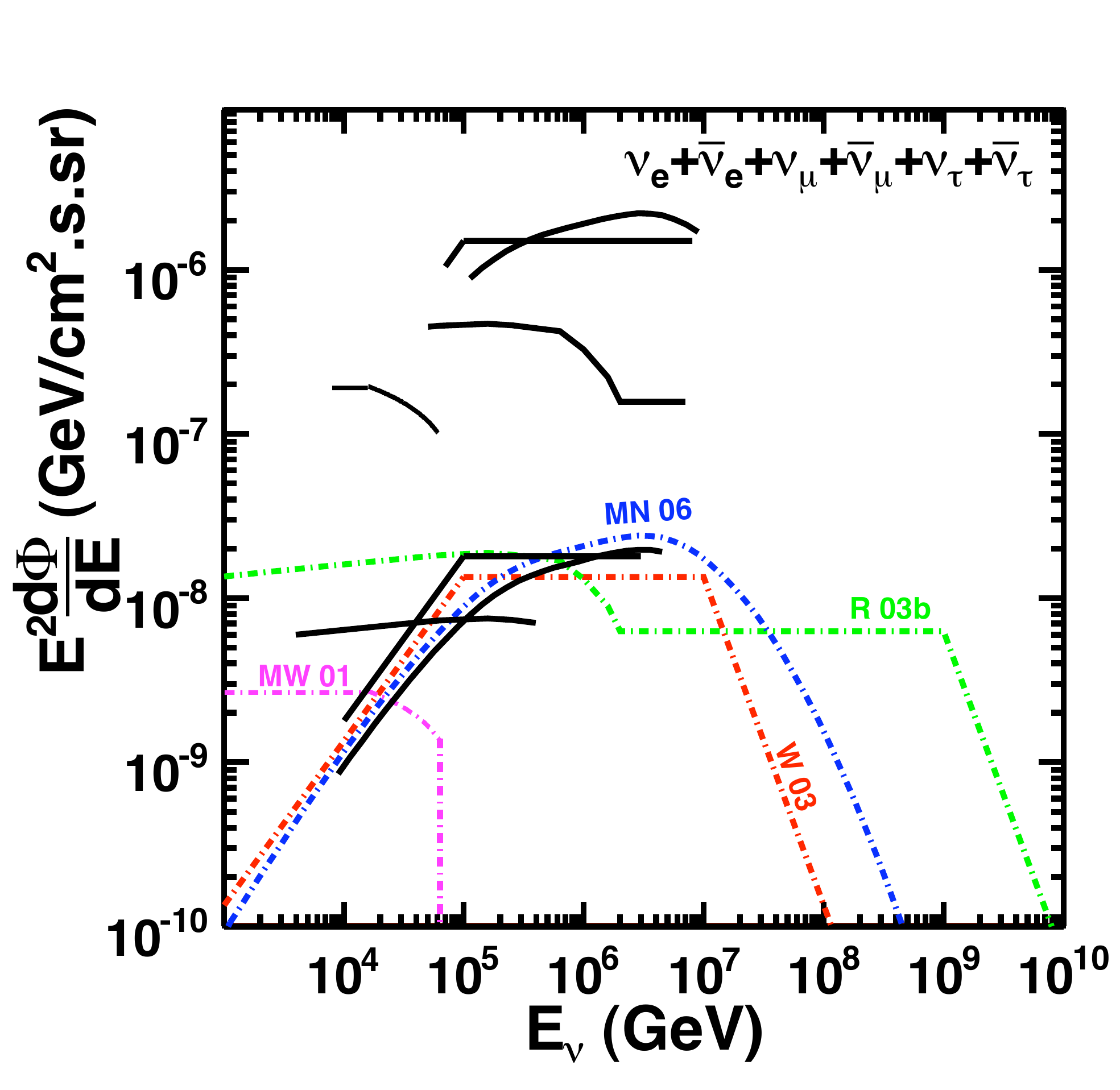}
    \caption{\label{fig:GRB}Integral limits on neutrino fluxes from
      GRB's, for several models of neutrino emission.}
  \end{minipage}\hspace{2pc}%
  \begin{minipage}{17pc}
    \includegraphics[width=17pc]{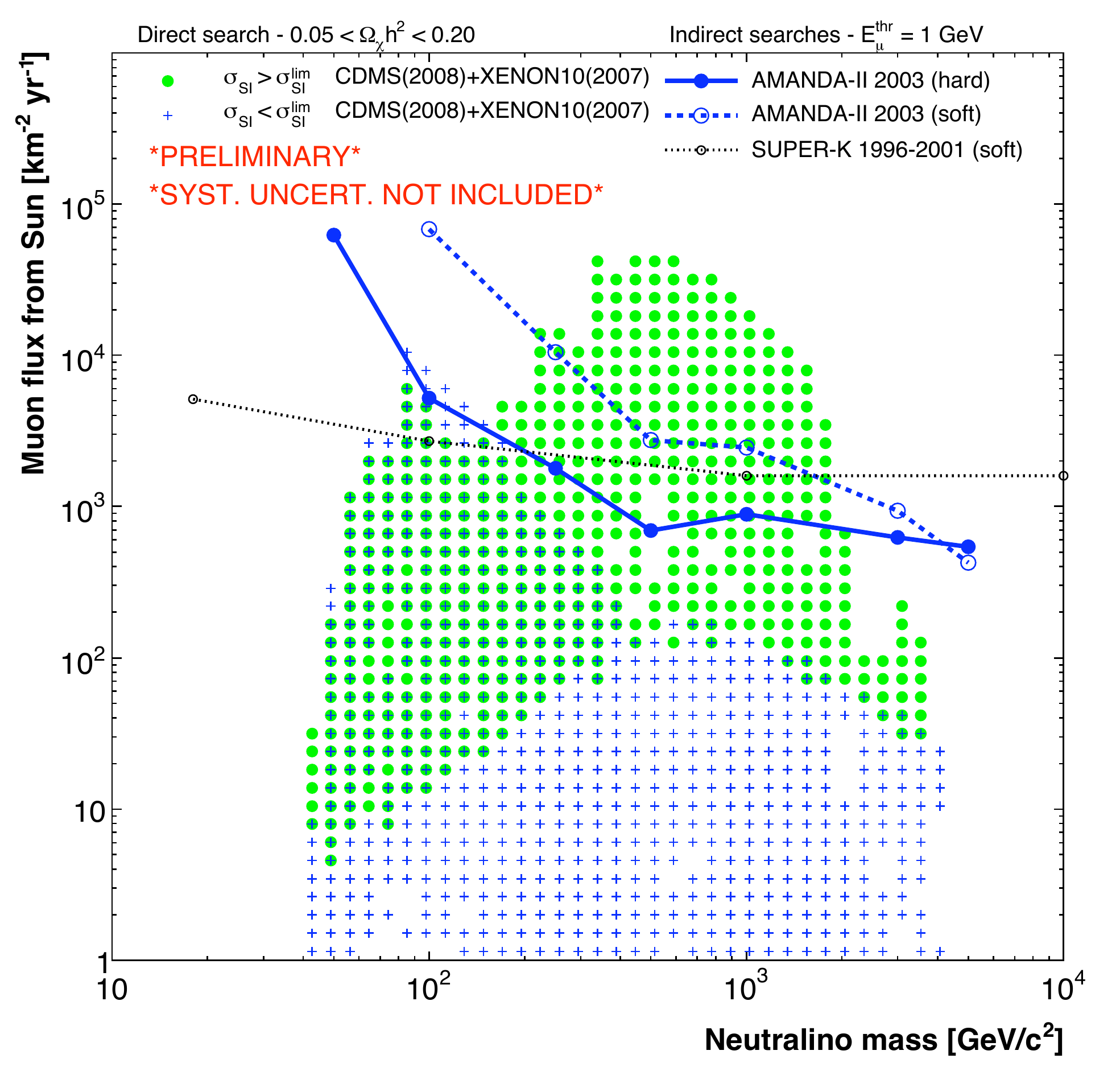}
    \caption{\label{fig:WIMP}Limits on the neutrino flux produced by
      neutralino annihilation in the Sun's gravitational well (details
      in text).}
  \end{minipage} 
\end{figure}

We have also searched for neutrino emission in conjunction with GRBs
observed by satellite instruments.  We have
searched for $\nu_\mu$ emission from over 400 GRBs observed in the
Northern Hemisphere over the
seven-year period \cite{Achterberg:2007nx}, with an additional search
conducted for an excess 
in cascade ($\nu_e$ CC and $\nu_x$ NC) events coincident with 73 observed
GRBs anywhere in the sky using the 2001--03 data set
\cite{Achterberg:2007qy}.  No such coincident events were 
observed, and limits have been placed on several models of GRB
emission
\cite{Waxman:1998yy,Waxman:2002wp,Meszaros:2001ms,Murase:2005hy,Razzaque:2002kb},
as shown in Fig.~\ref{fig:GRB}.  The limits
using the muon channel approach the predicted flux levels,
giving encouragement that IceCube will soon be able to probe neutrino
emission from GRBs.  

AMANDA is used to search for indirect evidence of dark matter that has
accumulated in the gravitational wells of the Earth
\cite{Achterberg:2006jf} and Sun and annihilate to produce neutrinos.
AMANDA's search is complementary to those conducted by direct
detection experiments because such experiments mainly constrain
models with spin-independent neutralino-nucleon scattering, whereas
neutralino capture in the Sun allows us to probe models where the
coupling is primarily spin-dependent \cite{Halzen:2005ar}.  Also,
direct and 
indirect searches probe different epochs of the history of the 
solar system and different parts of the WIMP velocity distribution. 
The limits from the AMANDA search using 2003 data are shown in
Fig.~\ref{fig:WIMP}, for neutralinos which annihilate through a hard
($\chi\chi \rightarrow W^+W^-$, or $\tau^+\tau^-$ at 50 GeV) or a soft
($\chi\chi \rightarrow b\bar{b}$) channel. Systematic uncertainties of
approximately 25\% are not included in the limits.
Each point on the plot represents one or more SUSY models which give
similar predicted neutralino-induced neutrino fluxes from the Sun; if
at least one model gives a scattering cross section above the limits from
direct detection experiments, a green point is plotted; if at least
one model predicts a cross section not excluded by direct detection
experiments, a blue cross is plotted.   

\section{Future Efforts}

Following seven years of operation as an independent detector, AMANDA
was incorporated into the growing IceCube neutrino telescope
\cite{icecube} during 
the 2006--07 austral summer.  AMANDA has operated as an IceCube
subdetector since then, with its higher density of instrumentation
augmenting IceCube's response to low energy events, which produce less
light than the TeV neutrinos for which IceCube is optimized.  With
IceCube strings deployed in the AMANDA volume improving the detector
response, and with outer IceCube strings providing an active veto
around that volume, the reach of the IceCube/AMANDA data will provide
improvements over the existing data set for topics such as the search
for dark matter.

\begin{figure}[h]
  \begin{minipage}{14pc}
    \includegraphics[width=14pc]{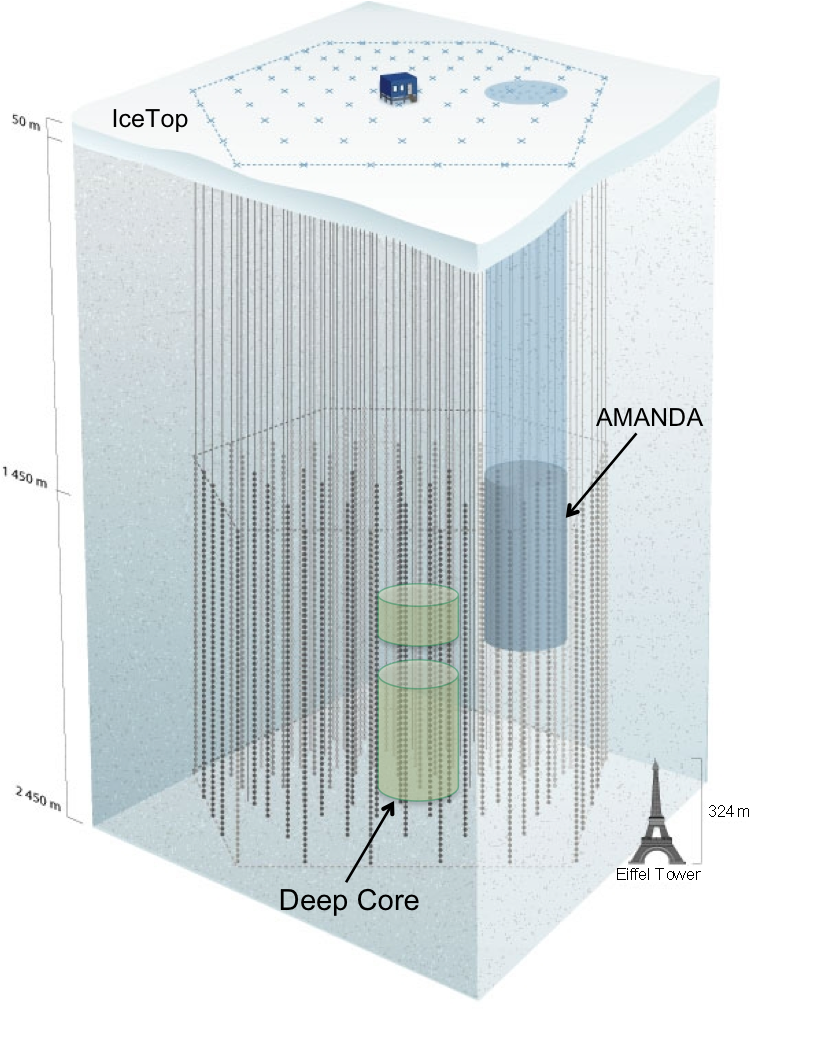}
    \caption{\label{fig:ICDC}Schematic of IceCube, including
      AMANDA and the new Deep Core detector.}
  \end{minipage}\hspace{2pc}%
  \begin{minipage}{22pc}
    \includegraphics[width=22pc]{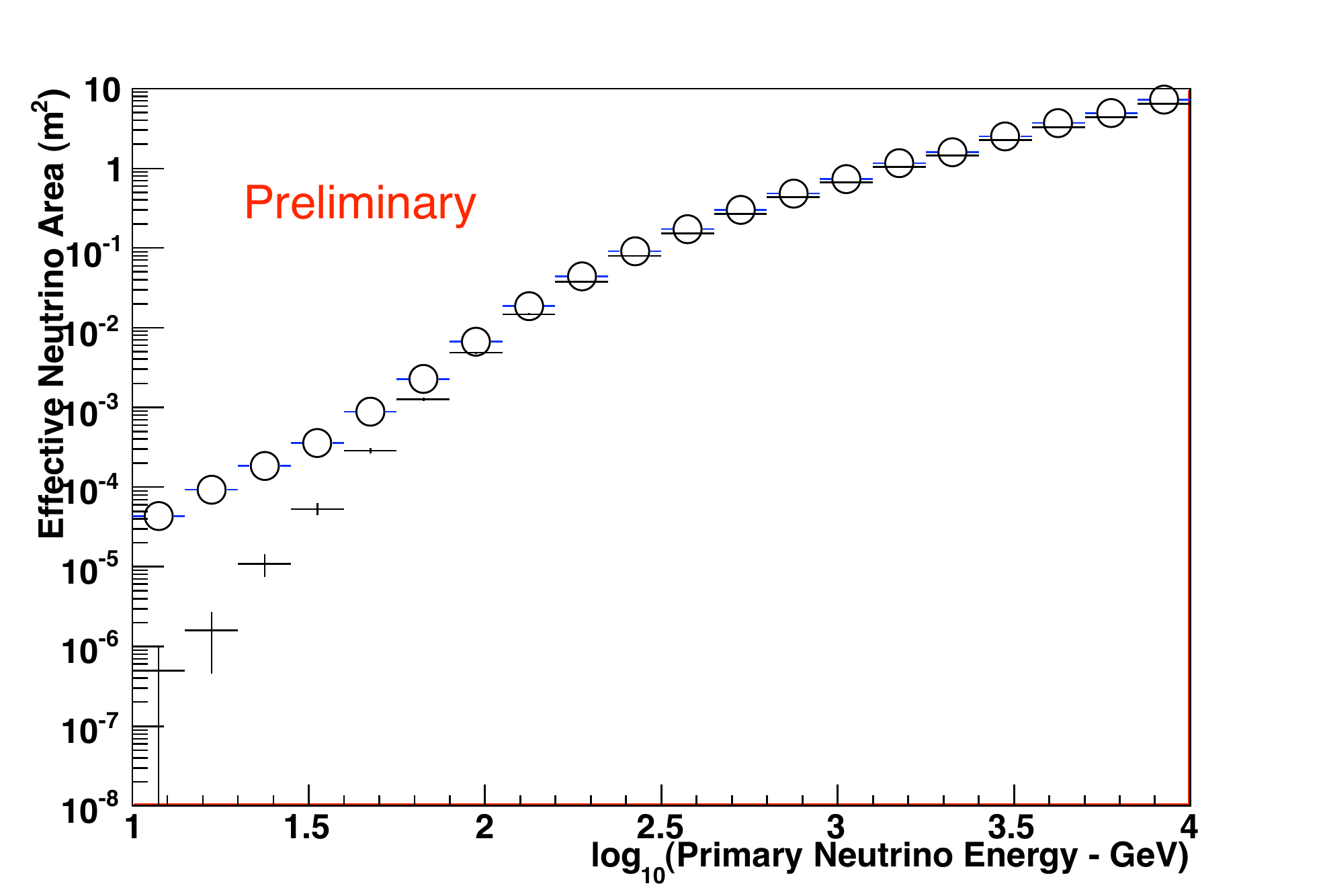}
    \caption{\label{fig:DCeffarea}Effective area to $\nu_\mu
      +\bar{\nu}_\mu$ for  
      IceCube with Deep Core at trigger level (circles), compared to
      IceCube alone (crosses).  At low energies, the improvement is
      more than an order of magnitude.} 
  \end{minipage} 
\end{figure}

Recently, funding has been approved to replace AMANDA with a new
Deep Core subdetector, shown in Figure~\ref{fig:ICDC}.  This
detector will be based on IceCube 
hardware, but with PMTs using a new photocathode material with 40\%
higher quantum efficiency.  Six new strings will be added, each with 50
OMs deployed between 2100 m and 2450 m and an additional 10 OMs
deployed at shallower depths to reinforce the veto shield against
downgoing cosmic ray muons.  The seven closest standard IceCube
strings will also be used as part of the fiducial Deep Core detector,
resulting in a volume comparable to AMANDA's.  The detector will be
located at the bottom 
center of IceCube, improving the veto efficiency for more horizontal
cosmic ray muons, and exploting the higher clarity of the ice at
depths below 2100 m.  Deep Core will increase the effective area of
IceCube significantly at energies below 100 GeV, as shown in
Fig.~\ref{fig:DCeffarea}. 

\end{document}